\documentclass[12pt]{article}  
\RequirePackage{fix-cm}
\RequirePackage{graphicx}
\usepackage{multicol}
\usepackage[normalem]{ulem}
\usepackage[english]{babel}
\usepackage{amssymb}
\usepackage{color}
\usepackage{rotating}
\usepackage{float}
\usepackage{rotating}
\usepackage{lscape}
\begin{document}

\title{AIC and BIC for cosmological interacting scenarios} 
\author{Fabiola Arevalo$^1$\footnote{fabiola.arevalo@ufrontera.cl}, Antonella Cid$^1$\footnote{acidm@ubiobio.cl} and Jorge Moya$^3$\footnote{jorgemoya@udec.cl}\\ \small{
$^1$ Departamento de Ciencias F\'isicas, Facultad de Ingenier\'ia y Ciencias, }\\
\small{Universidad de La Frontera, Temuco, Casilla 54-D, Chile }\\
\small{$^2$  Departamento de F{\'\i}sica and Grupo de Cosmolog{\'\i}a y Gravitaci\'on GCG-UBB, Universidad del B{\'\i}o-B{\'\i}o, }\\ \small{Casilla 5-C, Concepci\'on, Chile and} \\
 \small{
$^3$ Departamento de F{\'\i}sica, Universidad de Concepci\'on, }
\small{Casilla 160-C, Concepci\'on, Chile  }\\
}


%


\maketitle

\begin{abstract}
In this work we study linear and nonlinear cosmological interactions, which depend on dark matter and dark energy densities in the framework of General Relativity.
By using the Akaike information criterion (AIC) and the Bayesian information criterion (BIC) with data from SnIa (Union 2.1 and binned JLA), $\rm{H(z)}$, BAO and CMB we compare the interacting models among themselves and analyze whether more complex interacting models are favored by these criteria. In this context, we find some suitable interactions that alleviate the coincidence problem.
\end{abstract}

\section{Introduction} 

Since the seminal work of Riess and Perlmutter \cite{riessperl}, the astronomical observations of type Ia supernovae suggest that the late universe is in a phase of accelerated expansion driven by an unknown component dubbed dark energy. The fundamental nature of this late accelerated expansion remains unexplained, nevertheless recent observations \cite{Planck} are consistent with the simplest model, the $\Lambda$CDM scenario, which establishes that the energy density of the universe is dominated now by a non-relativistic fluid (dark matter) and a cosmological constant (dark energy). 

Despite the observational success of the $\Lambda$CDM scenario, this model has theoretical problems such as the fine-tuning problem and the coincidence problem \cite{Copeland} also there are some observational tensions recently reported, present when we use independently high redshift and low redshift data to constrain parameters \cite{tension}. Assuming that a departure of the $\Lambda$CDM scenario is needed, the simplest generalization is the so-called $\omega$CDM model, which describes dark energy as a perfect fluid with a constant state parameter $\omega$. Furthermore, models based on the interaction between dark matter and dark energy have been studied to describe the accelerated expansion. One of the first interacting models was proposed in Ref.\cite{ZPC}; it was mainly motivated to alleviate the coincidence problem in an interacting-quintessence scenario, focusing in an asymptotic attractor behavior for the ratio of the energy densities for the dark components. Since then, many interacting models with numerical and analytical solutions have emerged \cite{mimoso}-\cite{ruben}, including interactions with change of sign studied in Refs.\cite{wei}-\cite{schg}. A detailed review of cosmological interactions can be found in Ref.\cite{review} and some attempts to build an interaction from an action principle in Refs.\cite{new2}. In particular Refs.\cite{linear} present analytical solutions for a wide class of more elaborated interactions where the dark components are barotropic fluids with constant state parameters. Also, the question of how to discriminate among dark energy models (degeneracy problem \cite{Kunz}) has arisen in the context of interacting scenarios. In particular, there has been a debate on whether interacting models can be distinguished from modified dark energy equations of state, Chaplygin gas or modified gravity \cite{IntDeg}, which remains an open issue.

To compare different models of a certain physical phenomenon in light of the data there are criteria, based on Occam's razor (``among competing hypotheses, the one with the fewest assumptions should be selected''). These criteria measure the goodness of fitted models compared to a base model (see Refs.\cite{Trotta} and \cite{Liddle}). Two widely used criteria are the Akaike Information Criterion (AIC) \cite{ike} and the Bayesian Information Criterion (BIC) \cite{schwarz}. The first is an essentially frequentist criterion based on information theory and the second one follows from an approximation of the bayesian evidence valid for large sample size \cite{Trotta}.

In Cosmology AIC and BIC have been applied to discriminate cosmological models based on the penalization associated to the number of parameters that the model need to explain the data. Specifically, in Ref.\cite{Liddle2} the author performs cosmological model selection by using AIC and BIC in order to determinate the parameter set that better fit the WMAP3 data. Following this work in Ref.\cite{baye} the author considers more general models to the early universe description in light of AIC and BIC, also including the deviance information criterion. Regarding late universe description, the authors of Ref.\cite{Pol1} consider different models of dark energy and use information criteria to compare among them using the Gold sample of SnIa. Later on,  the authors of \cite{Castro} study interacting models, with an energy density ratio proportional to a power-law of the scale factor attempting to alleviate the coincidence problem. By using AIC and BIC, they compare the models among themselves and with $\Lambda$CDM considering data from SnIa, BAO and CMB. More recently, in Ref.\cite{baye2} the authors find that a particular interacting scenario is disfavored compared to $\Lambda$CDM. They study an interaction proportional to a power-law of the scale factor, by using AIC and BIC, and considering data from SnIa, $\rm{H(z)}$, BAO, Alcock--Paczynski test and CMB. 

In this work we analyze eight general types of interacting models with analytical  solution using Union 2.1 (or binned JLA)+$\rm{H(z)}$+BAO+CMB data under AIC and BIC. The main goal of our work is to investigate if complex interacting models are competitive in fitting the data and whether we could distinguish among them via the model comparison approach.

This paper is organized as follows: in section \ref{IM} we present and motivate eight types of interacting models with analytical solution to be revised. In section \ref{OA} we show the functions to be fitted and describe the information criteria to be used. In section \ref{AR} we present the analysis and results of the data fitting process and finally in section \ref{FR} we discuss our final remarks.

\section{Interacting Models}
\label{IM}

We work in the framework of general relativity by considering a spatially flat Friedmann-Lema\^itre-Robertson-Walker universe. The Friedmann equation is written as
\begin{equation}\label{claw}
3H^2=\rho,
\end{equation}
where $H=\dot{a}/a$ is the Hubble expansion rate, $a$ is the scale factor, the dot represents a derivative with respect to the cosmic time and we have considered $8\pi G=c=1$. From the energy-momentum tensor conservation we have
\begin{equation}\label{con}
\dot{\rho} +3H(\rho+p)=0,
\end{equation} 
where $\rho$ is the total energy density and $p$ is the effective pressure. First we consider that dark matter and dark energy are the relevant components of the total energy density at late times, i.e., $\rho=\rho_{\rm{x}}+\rho_{\rm{m}}$ and $p=p_{\rm{x}}+p_{\rm{m}}$ (where the subscripts $x$ and $m$ represent dark energy (DE) and  dark matter (DM), respectively). Furthermore, we consider a barotropic equation of state for both fluids, i.e., $p_{\rm{x}}=\omega_{\rm{x}}\rho_{\rm{x}}$ and $p_{\rm{m}}=\omega_{\rm{m}} \rho_{\rm{m}}$. 
To include a phenomenological interaction between these fluids, we separate the conservation Eq.(\ref{con}) into two equations
 \begin{eqnarray}
\dot{\rho}_{\rm{m}}+3\gamma_{\rm{m}} H\rho_{\rm{m}} &=&-Q, \label{rm}\\
\dot{\rho}_{\rm{x}}+3\gamma_{\rm{x}} H\rho_{\rm{x}} &=&Q,\label{rx}
 \end{eqnarray}
where $\gamma_{\rm{x}}=1+\omega_{\rm{x}}$, $\gamma_{\rm{m}}=1+\omega_{\rm{m}}$ and $Q$ represents the interaction function between dark matter and dark energy.
Using the change of variable $\eta=3\ln a$ and defining $()^\prime:=d/d\eta$, Eqs.(\ref{rm}) and (\ref{rx}) are rewritten as 
\begin{eqnarray}
\rho_{\rm{m}}^\prime+\gamma_{\rm{m}} \rho_{\rm{m}} &=&-\Gamma, \label{rm2}\\
\rho_{\rm{x}}^\prime+\gamma_{\rm{x}} \rho_{\rm{x}} &=&\Gamma, \label{rx2}
\end{eqnarray}
with $\Gamma=Q/3H$.  For $\Gamma > 0$ we have an energy transfer from DM to DE and for $\Gamma < 0$ we have the opposite energy transfer, from DE to DM. From Eqs.(\ref{rm2}) and (\ref{rx2}) and considering $\rho =\rho_{\rm{x}}+\rho_{\rm{m}}$ we can write $\rho_{\rm{x}}$ and $\rho_{\rm{m}}$ as \cite{linear}:
\begin{eqnarray}\label{rhoxm}
\rho_{\rm{x}}=\frac{\gamma_{\rm{m}}\rho +\rho^\prime}{\Delta},&& \rho_{\rm{m}}=-\frac{\gamma_{\rm{x}}\rho +\rho^\prime}{\Delta},
\end{eqnarray}
with $\Delta=\gamma_{\rm{m}}-\gamma_{\rm{x}}$
and from Eq.(\ref{con}) we get that
\begin{equation}
\label{peff}
p=-\rho-\rho^\prime.
\end{equation}
From Eqs.(\ref{rm2}) and (\ref{rhoxm}) we obtain the ``source equation'' defined in Ref.\cite{linear}:
\begin{equation}
\rho^{\prime \prime} + (\gamma_{\rm{x}} + \gamma_{\rm{m}})\rho^{\prime} + \gamma_{\rm{x}} \gamma_{\rm{m}} \rho =\Delta \Gamma, \label{source}
\end{equation}
valid for $\gamma_{\rm{x}}$ and $\gamma_{\rm{m}}$ constants.
We notice that due to (\ref{rhoxm}) every $\Gamma$ proportional to $\rho_{\rm{x}}$ and/or $\rho_{\rm{m}}$ in (\ref{source}) constitutes in fact, a differential equation for the variable $\rho$. Also, it is worth to mention that Eq.(\ref{source}) can be rewritten as a differential equation in terms of the deceleration parameter or in terms of a variable state parameter in a holographic context \cite{Fabiola}.

In this work we study eight types of interaction \cite{wei,linear}, defined as: $\Gamma_1=\alpha \rho_{\rm{m}}+\beta \rho_{\rm{x}}$,
$\Gamma_2=\alpha \rho_{\rm{m}}^\prime+\beta \rho_{\rm{x}}^\prime$, $\Gamma_3=\alpha \rho_{\rm{m}}\rho_{\rm{x}}/(\rho_{\rm{m}}+\rho_{\rm{x}})$,
$\Gamma_4=\alpha \rho_{\rm{m}}^2/(\rho_{\rm{m}}+\rho_{\rm{x}})$, $\Gamma_5=\alpha \rho_{\rm{x}}^2/(\rho_{\rm{m}}+\rho_{\rm{x}})$, $\Gamma_6=\alpha \rho$, $\Gamma_7=\alpha \rho^{\prime}$ and $\Gamma_8=\alpha q\rho=-\alpha(\rho+ 3\rho^{\prime}/2)$, where $q=-\left(1+\frac{\dot{H}}{H^2}\right)$ is the deceleration parameter, $\rho$ is the total energy density and $\alpha$, $\beta$ are constants. 

By rewriting Eq.(\ref{source}) as    
\begin{equation}\label{soursegeneral}
\rho\left[\rho^{\prime \prime}+b_1\rho^{\prime} +b_3\rho \right] + b_2\rho^{\prime 2}=0,
\end{equation}
it includes the eight types of interaction we are interested in, where the constants $b_1,b_2,b_3$ are different combinations of the relevant parameters depending on the particular interaction; see Table \ref{T1}. The general solution of Eq.(\ref{soursegeneral}) takes the form
\begin{equation}\label{10}
\rho(a)=\left[ C_1 a^{3\lambda_1} +C_2  a^{3\lambda_2}\right]^{\frac{1}{1+b_2}}.
\end{equation}
The integration constants in (\ref{10}) are given by
\begin{eqnarray}
C_1&=&-(3H_0^2)^{1+b_2}\left[\frac{\lambda_2 +\gamma_0(1+b_2)}{\lambda_1 -\lambda_2}\right], \nonumber \\
C_2&=&(3H_0^2)^{1+b_2}\left[\frac{\lambda_1 +\gamma_0(1+b_2)}{\lambda_1 -\lambda_2}\right],
\end{eqnarray}
and 
\begin{eqnarray}
\lambda_1 &=& -\frac{1}{2} \left(b_1 + \sqrt{b_1^2-4 b_3(1+b_2)}\right), \nonumber\\
\lambda_2 &=& -\frac{1}{2}\left(b_1 - \sqrt{b_1^2-4 b_3(1+b_2)}\right),\nonumber \\
\gamma_0  &=& \gamma_{\rm{m}}-\Omega_{ \rm{x}0}\Delta,
\end{eqnarray}
where $H_0$ and $\Omega_{ \rm{x}0}$ are the Hubble parameter and the value of the density parameter for DE today (i.e. $\Omega_{ \rm{x}0}=\rho_{ \rm{x}0}/3H_0^2$), respectively.

\begin{table}[ht!]
\centering
\caption{\label{T1} Definition of the constants $b_1,b_2$ and $b_3$ in terms of the relevant parameters for the studied interactions.}
\begin{tabular*}{\textwidth}{@{\extracolsep{\fill}}l l r l@{}}
\\ \hline
Interaction & $b_1$ & $b_2$ & $b_3$ \\ \hline 
$\Gamma_1=\alpha \rho_{\rm{m}}+\beta \rho_{\rm{x}}$ &$\gamma_{\rm{m}}+\gamma_{\rm{x}}+\alpha-\beta$ & $0$ &$\gamma_{\rm{m}}\gamma_{\rm{x}}+\alpha \gamma_{\rm{x}}-\beta\gamma_{\rm{m}}$ \\[1.5ex] 
$\Gamma_2=\alpha \rho_{\rm{m}}^\prime+\beta \rho_{\rm{x}}^\prime$ &$\displaystyle\frac{\gamma_{\rm_{m}}+\gamma_{\rm_{x}}+\alpha\gamma_{\rm_{x}}-\beta\gamma_{\rm_{m}}}{1+\alpha-\beta} $&$\displaystyle 0 $&$\displaystyle\frac{\gamma_{\rm{m}}\gamma_{\rm{x}}}{1+\alpha -\beta} $ \\[1.5ex] 
$\Gamma_3=\alpha \rho_{\rm{m}}\rho_{\rm{x}}/(\rho_{\rm{m}}+\rho_{\rm{x}})$ &$\displaystyle \gamma_{\rm{m}}+\gamma_{\rm{x}} +\alpha\frac{\gamma_{\rm{m}}+\gamma_{\rm{x}}}{\Delta}$&$\displaystyle \frac{\alpha}{\Delta}$&$\displaystyle \gamma_{\rm{m}}\gamma_{\rm{x}}+\alpha \frac{\gamma_{\rm{m}}\gamma_{\rm{x}}}{\Delta}$ \\[1.5ex] 
$\Gamma_4=\alpha \rho_{\rm{m}}^2/(\rho_{\rm{m}}+\rho_{\rm{x}})$ &$\displaystyle \gamma_{\rm{m}} +\gamma_{\rm{x}} -\frac{2\alpha\gamma_{\rm{x}}}{\Delta}$&$\displaystyle- \frac{\alpha}{\Delta}$&$\displaystyle \gamma_{\rm{m}}\gamma_{\rm{x}} -\frac{\alpha \gamma_{\rm{x}}^2}{\Delta} $ \\[1.5ex] 
$\Gamma_5=\alpha \rho_{\rm{x}}^2/(\rho_{\rm{m}}+\rho_{\rm{x}})$ &$\displaystyle \gamma_{\rm{m}} +\gamma_{\rm{x}} -\frac{2\alpha\gamma_{\rm{m}}}{\Delta}$&$\displaystyle -\frac{\alpha}{\Delta}$&$\displaystyle \gamma_{\rm{m}}\gamma_{\rm{x}} -\frac{\alpha \gamma_{\rm{m}}^2}{\Delta} $ \\[1.5ex] 
$\Gamma_6=\alpha \rho$ &$\gamma_{\rm{m}}+\gamma_{\rm{x}}$ & $0$ &$\gamma_{\rm{m}}\gamma_{\rm{x}}-\alpha\Delta$ \\[1.5ex] 
$\Gamma_7=\alpha \rho^\prime$ &$\displaystyle \gamma_{\rm_{m}}+\gamma_{\rm_{x}}-\alpha\Delta $&$\displaystyle 0 $&$\displaystyle\gamma_{\rm{m}}\gamma_{\rm{x}}$ \\[1.5ex] 
$\Gamma_8=\alpha q\rho=-\alpha(\rho+\frac{3}{2}\rho^\prime)$ &$\displaystyle\gamma_{\rm_{m}}+\gamma_{\rm_{x}}+\frac{3}{2}\alpha\Delta $&$\displaystyle 0 $&$\displaystyle\gamma_{\rm{m}}\gamma_{\rm{x}}+\alpha\Delta $ \\  \hline
\end{tabular*}
\end{table}
The nature of cosmic interaction remains unknown, however, physical motivation to study most of the interactions in Table \ref{T1} can be found in the literature. These interactions are worth to study because it has been shown that most of them could alleviate the coincidence problem \cite{linear,Arevalo:2011hh}. It was demonstrated in Ref.\cite{0712.0565} that an interaction proportional to $H\rho_x$ could be consistent with the second law of thermodynamics if the energy transfer is from DE to DM, also, in Ref.\cite{1507.00187} it was shown that interactions proportional to $H(\rho_m+\rho_x)$ or $H\rho_m$ can arise by imposing simple thermodynamic arguments based on the evolution of the ratio $\rho_m/\rho_x$. For interactions proportional to $\rho_m'$, $\rho_x'$ or a linear combination of both, we note from Eqs.(\ref{rm}) and (\ref{rx}), that these interactions can be rewritten in terms of interactions proportional to a linear combination of $\rho_m$ and $\rho_x$. We can find a physical motivation to nonlinear interactions in Ref.\cite{0901.1215}, in the context of holographic interacting models. On the other hand, sign-changeable interaction was found to be preferred by the data in Refs.\cite{cai}-\cite{schg}. It has also been shown that a late-time interaction can alleviate the tension that arises in $\Lambda$CDM between the Hubble constant measurements from Planck and the Hubble Space Telescope \cite{1406.7297}. In Refs.\cite{new3} it was shown that interaction proportional to $H\rho_m$, $H\rho_x$ and $H\rho_m\rho_x/(\rho_m+\rho_x)$ can have stable cosmological perturbations during the whole expansion history, i.e. these interactions could consistently describe the linear evolution of growing structures, without large-scale instabilities.

On the other hand, the effective energy density (\ref{10}) associated to the general solution of our interactions has an effective pressure (\ref{peff}) corresponding to a variable modified Chaplygin gas \cite{Chaplygin} given by

\begin{equation}
\label{vmChG}
p= -\rho\left(1+\frac{\lambda_1}{1+b_2}\right) -C_2 \frac{\lambda_2 -\lambda_1}{1+b_2}\rho^{-b_2}a^{3\lambda_2}.
\end{equation}

This means that the considered interactions can be interpreted as a single fluid model in a unified description of the dark sector inherently.

Also, the effective energy density (\ref{10}) can be interpreted as a non-interacting description of the dark sector with a variable barotropic index for the dark energy component given by
\begin{eqnarray}
\label{vbi}
\gamma_x(a)=-\frac{C_1 \lambda_1 a^{3\lambda_1}+C_2 \lambda_2 a^{3\lambda_2}+\gamma_m \rho_{m0} a^{-3\gamma_m}}{\left[C_1 a^{3\lambda_1}+C_2a^{3\lambda_2}-\rho_{m0} a^{-3\gamma_m}+\rho_{x0}\right]},
\end{eqnarray}
where $\rho_{m0}$ and $\rho_{x0}$ are, respectively, the current values of the DM and DE densities.
The inverse approach has been considered in Ref.\cite{1505.04443}, where the relation between a given variable state parameter and a reconstructed interaction has been addressed using Gaussian processes.

The solution in Eq.(\ref{10}) is valid for late-time evolution, nevertheless if we are interested in data from BAO and/or CMB, which consider high redshifts, we need to take into account the radiation contribution in the equations as well as the baryons contribution. If we consider from here on $\rho=\rho_{\rm{m}}+\rho_{\rm{x}}+\rho_{\rm{r}}+\rho_{\rm{b}}$, with $\rho_{\rm{r}}$ the energy density of relativistic matter and $\rho_{\rm{b}}$ the energy density of baryons, which we assume are non-interacting with the dark fluids, then the solution of Eq.(\ref{soursegeneral}) is given by
\begin{eqnarray}
\rho(a)&=\left[ C_1 a^{3\lambda_1} +C_2  a^{3\lambda_2}\right]^{\frac{1}{1+b_2}} +3H_0^2\left(\frac{\Omega_{\rm{r0}}}{a^{4}}+\frac{\Omega_{\rm{b0}}}{a^{3}}\right),\label{solu}
\end{eqnarray} 
where $\Omega_{\rm{r0}}$ and $\Omega_{\rm{b0}}$ are the current values of the density parameters for radiation and baryons, respectively, and the constants $C_1$ and $C_2$ (for interactions $\Gamma_1$ to $\Gamma_5$) are modified to 
\begin{eqnarray}
C_1&=& \left[3H_0^2(\Omega_{\rm{x0}}+\Omega_{\rm{m0}})\right]^{1+b_2} -C_2,  \nonumber \\
C_2&=&- \frac{(3H_0^2)^{1+b_2}\left[(\Omega_{\rm{x0}}\gamma_{\rm{x}}+\Omega_{\rm{m0}}\gamma_{\rm{m}})(1+b_2) \right]}{(\Omega_{\rm{x0}}+\Omega_{\rm{m0}})^{-b_{2}}(\lambda_2-\lambda_1)}
-\frac{(3H_0^2)^{1+b_2}\lambda_1}{(\lambda_2-\lambda_1)(\Omega_{\rm{x0}}+\Omega_{\rm{m0}})^{-1-b_{2}}}.
\end{eqnarray}

The values of $b_1,b_2,b_3$ are the same for both cases, including radiation and baryons or not;  see Table 1.

For interactions $\Gamma_6 - \Gamma_8$ we can decompose the general solution into a homogeneous solution $\rho_{\rm{h}}$ and a particular solution $\rho_{\rm{p}}$, then the general solution is given by $\rho=\rho_{\rm{h}}+\rho_{\rm{p}}$. The homogeneous part of the solution $\rho_{\rm{h}}$ corresponds to (\ref{solu}) and the particular solution is given by  
\begin{equation}\label{solup}
\rho_{\rm{p}}(a)=-9M_{ri}a^{-4}-M_{bi}a^{-3},
\end{equation}  
where $M_{ri}=-3H_0^2\delta_{ri}\Omega_{\rm{r}0}\Delta /(12b_1 - 9 b_3 -16)$, $M_{bi}=3H_0^2\delta_{bi}\Omega_{\rm{b}0}\Delta /(2b_1 - 2 b_3 -2)$, $(\delta_{r6},\delta_{r7},\delta_{r8})=\left(-\alpha,\frac{4}{3}\alpha,-\alpha\right)$, $(\delta_{b6},\delta_{b7},\delta_{b8})=\left(2\alpha,-2\alpha,\alpha\right)$ and now the constants $C_1$ and $C_2$ are given by
\begin{eqnarray}\label{ces1}
C_1&=&3H_0^2(\Omega_{\rm{x0}}+\Omega_{\rm{m0}})+9M_{ri}+M_{bi} -C_2,\\
C_2&=& \frac{3H_0^2\Delta \Omega_{\rm{x0}} }{\lambda_2-\lambda_1}-\frac{(9\lambda_1 +12)M_{ri} }{\lambda_2-\lambda_1}-\frac{(\lambda_1 +1)M_{bi} }{\lambda_2-\lambda_1}
-\frac{3H_0^2(\Omega_{\rm{x0}}+\Omega_{\rm{m0}})(\gamma_{\rm{m}}+\lambda_1)}{\lambda_2-\lambda_1}.
\label{ces2}
\end{eqnarray}

Additionally, to examine the coincidence problem we use the coincidence parameter $r$ defined as
\begin{equation}\label{r}
r=\frac{\rho_{\rm{m}}}{\rho_{\rm{x}}}.
\end{equation}
We can therefore calculate the asymptotic limit of  $r(a)$ when $a$ tends to $\infty$. For all our interactions we get
\begin{equation}
r_{\infty}=-\left[1+ \frac{2(\gamma_{\rm{x}}-1)(1+b_2)}{2(1+b_2)-b_1+\sqrt{b_1^2-4b_3(1+b_2)}}\right],
\end{equation}  
a constant that depends on the state parameters and interaction parameters. The author of Ref.\cite{linear} noticed that, for a constant and positive $\gamma_x$ and for an interacting term proportional to $\rho$, $\rho'$ or $\rho_x$, there is obtained a positive $r$ parameter asymptotically constant, alleviating in this sense the coincidence problem.
Furthermore, the authors in Ref. \cite{Arevalo:2011hh}, analyze nonlinear models $\Gamma_3$, $\Gamma_4$ and $\Gamma_5$,  concluding that the last two interactions may alleviate the coincidence problem also.

In this section we have assumed that an interacting scenario of DM and DE can be described in terms of fluids with a constant state parameter. In this sense, the source equation (\ref{source}) allows us to study a family of interacting scenarios recast in a single functional form (\ref{10}), where we have considered the more common linear and nonlinear interactions and also a naturally sign-changeable interaction. Besides, these interactions can be interpreted, at the background level, in terms of a unified fluid description with a variable modified Chaplygin gas (\ref{vmChG}) or, in terms of a variable equation of state (\ref{vbi}) for the dark energy component with a non-interacting dark sector.

\section{Observational analysis and model selection}
\label{OA}
In order to constrain the interacting models, we use the following data:
i) distance modulus of type Ia supernovae from: 580 data points from the Union 2.1 compilation \cite{SN21} or 31 data points of binned data from the JLA compilation \cite{JLA}, ii) 28 data points from $\rm{H(z)}$ data \cite{liao}.
iii) For BAO data we use: the acoustic parameter (3 data points from the WiggleZ experiment \cite{blake}) and the distance ratio (2 data points from the  SDSS \cite{percival} and 1 data point from the 6dFGS surveys \cite{beutler}). From CMB data we consider the position of the first peak in the CMB anisotropy spectrum \cite{Planck2}.

To fit the cosmological models to the data we use the Chi-square method. Each dataset (SnIa, $\rm{H(z)}$, WiggleZ, SDSS, 6dFGS and CMB)  has a corresponding Chi-square function ($\chi^2_{\rm{Sn}}$, $\chi^2_{\rm{H(z)}}$, $\chi^2_{\rm{WiggleZ}}$, $\chi^2_{\rm{SDSS}}$, $\chi^2_{\rm{6dFGS}}$, $\chi^2_{\rm{CMB}}$) which is used to calculate the overall $\chi^2$ function. These functions are defined according to each dataset.

For SnIa we have the $\chi^2$ function defined as
\begin{equation}
\chi^2_{\rm{Sn}}=\sum_{i=1}^{N_{Sn}}\frac{(\mu_{i,\rm{th}} -\mu_{i,\rm{obs}})^2}{\sigma^2_{\mu_i}},
\end{equation}
where $\mu$ is the distance modulus defined in appendix (\ref{mu}), ``th'' represents the theoretical function, ``obs'' the observed value, $\sigma_{\mu_i}$ is the uncertainty associated to the observed value and $N_{Sn}$ is the data number of SnIa in the compilation of Union 2.1 or the number of binned data for the JLA compilation.
Similarly, for $\rm{H(z)}$ we have the $\chi^2$ function for the Hubble expansion rate (\ref{Hz}):
\begin{equation}
\chi^2_{\rm{H(z)}}=\sum_{i=1}^{N_H}\frac{(H_{i,\rm{th}} -H_{i,\rm{obs}})^2}{\sigma^2_{H_i}},
\end{equation} 
where $N_{H}$ is the data number of $\rm{H(z)}$ data.

For BAO's measurements we have $\chi^2_{\rm{BAO}}$ given by
\begin{equation}
\chi^2_{\rm{BAO}}= \chi^2_{\rm{WiggleZ}}+\chi^2_{\rm{SDSS}}+\chi^2_{\rm{6dFGS}}.
\end{equation}

In the case of WiggleZ we use the inverse of the covariance matrix $C^{-1}_{\rm{WiggleZ}}$ \cite{blake},
\begin{equation}
\chi^2_{\rm{WiggleZ}}=(A_{\rm{th}}-A_{\rm{obs}}) C^{-1}_{\rm{WiggleZ}} (A_{\rm{th}}-A_{\rm{obs}})^{T},
\end{equation}
where $A_{\rm{th}}$ is the theoretical acoustic parameter defined in the appendix (\ref{acoustic}), the observational values of this parameter are given by $A_{\rm{obs}}=(0.474,0.442,0.424)$ at redshifts $z=(0.44,0.6,0.73)$, respectively, and 
\begin{equation}
C^{-1}_{\rm{WiggleZ}}=\left(
\begin{array}{ccc}
1040.3 &-807.5 &336.8 \\
-807.5 &3720.3 &-1551.9\\
 336.8&-1551.9 &2914.9
\end{array}
 \right).
\end{equation}
Analogously, for SDSS \cite{percival} we have 
\begin{equation}
\chi^2_{\rm{SDSS}}=(d_{\rm{th}}-d_{\rm{obs}}) C^{-1}_{\rm{SDSS}} (d_{\rm{th}}-d_{\rm{obs}})^{T},
\end{equation}
where $d_{\rm{th}}$ is the theoretical distance ratio defined in the appendix, see Eq.(\ref{dr}), the observational values are given by $d_{\rm{obs}}=(0.1905,0.1097)$ at redshifts $z=(0.2,0.35)$ and the inverse of the covariance matrix is
\begin{equation}
C^{-1}_{\rm{SDSS}}=\left(
\begin{array}{cc}
30124 &-17227 \\
-17227 &86977
\end{array}
 \right).
\end{equation}
The data point of the 6dFGS is given by
\begin{equation}
\chi^2_{\rm{6dFGS}}= \left(\frac{d_{\rm{th}}-d_{\rm{obs}}}{\sigma_d}\right)^2,
\end{equation}
with the observed distance ratio $d_{\rm{obs}}=0.336$ and $\sigma_d=0.015$, at redshift $z=0.106$ \cite{beutler}. 

Finally, we consider the position of the first peak of the CMB anisotropy as a background data coming from early universe's physics. It is common to consider also the shift parameter, but the derivation of this parameter is assuming a $\Lambda$CDM scenario today \cite{0702343}. It is more consistent to consider only the position of the first peak to test interacting models because it only depends on pre-recombination physics (see the discussion in Refs. \cite{Saulo}) and in this sense, it can be considered in our work as a good approximation. The $\chi^2$ contribution of the position of the first peak $l_1$ is given by
\begin{equation}
\chi^2_{\rm{CMB}}= \left(\frac{l_{1\rm{th}}-l_{1\rm{obs}}}{\sigma_l}\right)^2,
\end{equation}
where $l_{1\rm{th}}$ is the position of the first peak defined in the appendix (\ref{l1}), $l_{1\rm{obs}}$ is the observed position of the first peak, $l_{1\rm{obs}}=220.0$ and $\sigma_l=0.5$ \cite{Planck2}.

In order to find the best fit model parameters we perform a joint analysis using all the data, we minimize the overall $\chi^2$ function defined as
\begin{equation}
\chi^2=\chi^2_{\rm{Sn}}+\chi^2_{\rm{H(z)}}+\chi^2_{\rm{BAO}}+\chi^2_{\rm{CMB}}.
\end{equation}

Each Chi-squared function depends on the parameters of the model. Based on statistical analysis we can determine which models are ``better'' taking into account how many parameters do the models need and how well do they fit the data. In this work we use two criteria, the Akaike Information Criterion (AIC) and the Bayesian Information Criterion (BIC).
The AIC parameter is defined through the relation \cite{ike}:
\begin{equation}
\mbox{AIC}= \chi^2_{\rm{min}}+2d,
\end{equation}
where $d$ is the number of free parameters in the model and $\chi^2_{\rm{min}}$ is the minimum value of the $\chi^2$ function.
The ``preferred model'' for this criterion is the one with the smaller value of AIC. This criterion  ``penalizes'' models according to the number of free parameters that they have.

To compare the model $k$ with the model $l$, we calculate $\Delta\mbox{AIC}_{kl}=\mbox{AIC}_{k}-\mbox{AIC}_{l}$, which can be interpreted as ``evidence in favor'' of the model $k$ compared to the model $l$. For $0\leq\Delta\mbox{AIC}_{kl}<2$ we have ``strong evidence in favor'' of model $k$, for $4<\Delta\mbox{AIC}_{kl}\leq7$ there is ``little evidence in favor'' of the model $k$, and for  $\Delta\mbox{AIC}_{kl}>10$ there is basically ``no evidence in favor'' of model $k$ \cite{baye}.

On the other hand, the Bayesian criterion is defined through the relation
\begin{equation}
\mbox{BIC}= \chi^2_{\rm{min}}+d\ln N,
\end{equation}
where $N$ is the number of data points.
Similarly to $\Delta\mbox{AIC}_{kl}$, $\Delta\mbox{BIC}_{ij}=\mbox{BIC}_{i}-\mbox{BIC}_{j}$ can be interpreted as ``evidence against'' the model $i$ compared to the model $j$. For $0\leq\Delta\mbox{BIC}_{ij}<2$ there is ``not enough  evidence against'' the model $i$, for $2\leq\Delta\mbox{BIC}_{ij}<6$ there is ``evidence against'' the model $i$ and for $6\leq\Delta\mbox{BIC}_{ij}<10$ there is ``strong  evidence against'' model $i$ \cite{baye}.

\section{Analysis and results}
\label{AR}

For model fitting we use the Chi-Square method with the Levenberg--Marquardt algorithm implemented in the package lmfit of Python.\footnote{https://www.python.org}
For all the studied interactions we consider a fixed $\gamma_{\rm{m}}$. The search ranges of the free parameters in our models are: $\Omega_{\rm{m}}\in[0,1]$, $\gamma_{\rm{x}}\in[-0.5,0.5]$, $\alpha \in [-0.5,0.5]$, $\beta \in [-0.5,0.5]$ and $h \in [0,1]$. We use the combined datasets Union 2.1 (or binned JLA), $\rm{H(z)}$, BAO and CMB for the data fitting and we restrict our analysis to a maximum of four free parameters for each model.

We consider two possible scenarios, one where we fix parameters such as $\gamma_{\rm{m}}=1$ which corresponds to a cold dark matter scenario or we fix $\gamma_{\rm{m}}=1$ and $\gamma_{\rm{x}}=0$ that corresponds to a $\Lambda(t)$CDM model \cite{L(t)}. For these scenarios we can additionally fix the parameters associated with different models of phenomenological interaction, $\alpha$ and/or $\beta$.
 
\begin{sidewaystable}
\begin{center}
\caption{Results of the data fitting using the joint analysis from Union 2.1, $\rm{H(z)}$, BAO and CMB. The error informed corresponds to 68$\%$ confidence level. Fixed means that the parameter was set to zero and the dashed lines mean that the model does not have that parameter. The derived parameters are: the current value of the deceleration parameter $q_0$, the value of the effective state parameter today $w_{\rm{eff}}$ and the calculated age of the universe in Gy. The AIC and BIC parameters are indicated in each case.
\label{Tablea}}
\scalebox{0.65}{
\begin{tabular}{llllllllllll}
\\ \hline
Model&$\Omega_{\rm{m}0}$ & $\gamma_{\rm{x}}$ & $\alpha$ & $\beta$ &  $h$ & $q_0$ & $\omega_{\rm{eff}}$ & Age & AIC & BIC \\ \hline
$\Gamma_{1a}$ &$0.239\pm 0.021$&Fixed&$0.0004\pm 0.0028$&$0.0060\pm 0.0423$&$0.699\pm 0.003$&$-0.573\pm 0.031$&$-0.716\pm 0.021$&$13.633\pm 0.410$&$588.450$&$606.137$\\ [1.0ex]
$\Gamma_{1b}$ &$0.247\pm 0.027$&$-0.059\pm 0.099$&Fixed&$0.0045\pm 0.0044$&$0.701\pm 0.004$&$-0.624\pm 0.114$&$-0.749\pm 0.076$&$13.616\pm 0.420$&$587.422$&$605.108$\\ [1.0ex]
$\Gamma_{1c}$ &$0.250\pm 0.016$&$-0.061\pm 0.058$&$0.0000\pm 0.0049$&Fixed&$0.701\pm 0.004$&$-0.622\pm 0.066$&$-0.748\pm 0.044$&$13.618\pm 0.249$&$587.436$&$605.123$\\ [1.0ex]
$\Gamma_{1d}$ &$0.250\pm 0.022$&$-0.060\pm 0.086$&$0.0000\pm 0.0020$&$0.0000\pm 0.0020$&$0.701\pm 0.004$&$-0.621\pm 0.098$&$-0.747\pm 0.065$&$13.618\pm 0.347$&$587.435$&$605.122$\\ [1.0ex]
$\Gamma_{1e}$ &$0.239\pm 0.014$&Fixed&Fixed&$0.0010\pm 0.0006$&$0.699\pm 0.003$&$-0.573\pm 0.021$&$-0.715\pm 0.014$&$13.672\pm 0.191$&$586.502$&$599.767$\\ [1.0ex]
$\Gamma_{1f}$ &$0.241\pm 0.014$&Fixed&$0.0003\pm 0.0001$&Fixed&$0.699\pm 0.003$&$-0.570\pm 0.021$&$-0.713\pm 0.014$&$13.650\pm 0.190$&$586.464$&$599.729$\\ [1.0ex]
$\Gamma_{1g}$ &$0.241\pm 0.013$&Fixed&$0.0003\pm 0.0015$&$0.0003\pm 0.0015$&$0.699\pm 0.003$&$-0.570\pm 0.019$&$-0.713\pm 0.013$&$13.647\pm 0.174$&$586.462$&$599.726$\\ [1.0ex]
$\Gamma_{2a}$ &$0.238\pm 0.015$&Fixed&$0.0004\pm 0.0053$&$-0.2402\pm 52.2394$&$0.699\pm 0.003$&$-0.575\pm 0.022$&$-0.717\pm 0.015$&$13.695\pm 0.232$&$588.666$&$606.352$\\ [1.0ex]
$\Gamma_{2b}$ &$0.249\pm 0.031$&$-0.058\pm 0.067$&Fixed&$0.0033\pm 0.8906$&$0.701\pm 0.004$&$-0.620\pm 0.086$&$-0.747\pm 0.058$&$13.620\pm 0.570$&$587.434$&$605.120$\\ [1.0ex]
$\Gamma_{2c}$ &$0.251\pm 0.019$&$-0.064\pm 0.071$&$-0.0001\pm 0.0026$&Fixed&$0.701\pm 0.004$&$-0.624\pm 0.081$&$-0.749\pm 0.054$&$13.605\pm 0.293$&$587.456$&$605.143$\\ [1.0ex]
$\Gamma_{2d}$ &$0.247\pm 0.020$&$-0.054\pm 0.073$&$0.0003\pm 0.0038$&$0.0003\pm 0.0038$&$0.701\pm 0.004$&$-0.619\pm 0.084$&$-0.746\pm 0.056$&$13.648\pm 0.314$&$587.489$&$605.175$\\ [1.0ex]
$\Gamma_{2f}$ &$0.242\pm 0.014$&Fixed&$-0.0003\pm 0.0001$&Fixed&$0.699\pm 0.003$&$-0.569\pm 0.021$&$-0.713\pm 0.014$&$13.642\pm 0.190$&$586.467$&$599.732$\\ [1.0ex]
$\Gamma_{3}$ &$0.251\pm 0.024$&$-0.030\pm 0.077$&$0.0005\pm 0.0017$&$---$&$0.700\pm 0.004$&$-0.587\pm 0.090$&$-0.725\pm 0.060$&$13.561\pm 0.359$&$588.322$&$606.009$\\ [1.0ex]
$\Gamma_{3a}$ &$0.245\pm 0.014$&Fixed&$0.0004\pm 0.0001$&$---$&$0.698\pm 0.003$&$-0.564\pm 0.021$&$-0.710\pm 0.014$&$13.602\pm 0.191$&$586.625$&$599.890$\\ [1.0ex]
$\Gamma_{4}$ &$0.254\pm 0.017$&$-0.068\pm 0.067$&$0.0005\pm 0.0023$&$---$&$0.701\pm 0.004$&$-0.622\pm 0.076$&$-0.748\pm 0.050$&$13.568\pm 0.269$&$587.692$&$605.379$\\ [1.0ex]
$\Gamma_{4a}$ &$0.240\pm 0.014$&Fixed&$0.0001\pm 0.0001$&$---$&$0.699\pm 0.003$&$-0.571\pm 0.021$&$-0.714\pm 0.014$&$13.662\pm 0.189$&$586.476$&$599.741$\\ [1.0ex]
$\Gamma_{5}$ &$0.250\pm 0.023$&$-0.059\pm 0.084$&$-0.0040\pm 0.0038$&$---$&$0.701\pm 0.004$&$-0.619\pm 0.096$&$-0.746\pm 0.064$&$13.622\pm 0.357$&$587.449$&$605.135$\\ [1.0ex]
$\Gamma_{5a}$ &$0.235\pm 0.013$&Fixed&$0.0158\pm 0.0049$&$---$&$0.699\pm 0.003$&$-0.580\pm 0.019$&$-0.720\pm 0.013$&$13.664\pm 0.177$&$586.620$&$599.885$\\ [1.0ex]
$\Gamma_{6}$ &$0.243\pm 0.018$&$-0.044\pm 0.066$&$0.0016\pm 0.0017$&$---$&$0.701\pm 0.004$&$-0.615\pm 0.076$&$-0.743\pm 0.051$&$13.647\pm 0.290$&$587.440$&$605.126$\\ [1.0ex]
$\Gamma_{6a}$ &$0.236\pm 0.014$&Fixed&$0.0019\pm 0.0009$&$---$&$0.699\pm 0.003$&$-0.577\pm 0.021$&$-0.718\pm 0.014$&$13.672\pm 0.196$&$586.119$&$599.384$\\ [1.0ex]
$\Gamma_{7}$ &$0.244\pm 0.019$&$-0.046\pm 0.046$&$-0.0016\pm 0.0079$&$---$&$0.701\pm 0.003$&$-0.616\pm 0.058$&$-0.744\pm 0.039$&$13.651\pm 0.292$&$587.421$&$605.107$\\ [1.0ex]
$\Gamma_{7a}$ &$0.237\pm 0.013$&Fixed&$-0.0018\pm 0.0006$&$---$&$0.699\pm 0.003$&$-0.576\pm 0.020$&$-0.717\pm 0.013$&$13.677\pm 0.184$&$586.103$&$599.367$\\ [1.0ex]
$\Gamma_{8}$ &$0.230\pm 0.018$&$-0.018\pm 0.063$&$0.0012\pm 0.0021$&$---$&$0.701\pm 0.004$&$-0.606\pm 0.074$&$-0.738\pm 0.049$&$13.805\pm 0.299$&$589.305$&$606.991$\\ [1.0ex]
$\Gamma_{8a}$ &$0.239\pm 0.014$&Fixed&$0.0034\pm 0.0013$&$---$&$0.699\pm 0.003$&$-0.573\pm 0.020$&$-0.715\pm 0.014$&$13.679\pm 0.190$&$586.097$&$599.362$\\ [1.0ex]
$\omega$CDM &$0.249\pm 0.016$&$-0.059\pm 0.081$&$---$&$---$&$0.701\pm 0.004$&$-0.621\pm 0.090$&$-0.747\pm 0.060$&$13.620\pm 0.274$&$585.435$&$598.700$\\ [1.0ex]
$\Lambda$CDM &$0.239\pm 0.007$&$---$&$---$&$---$&$0.699\pm 0.003$&$-0.572\pm 0.010$&$-0.715\pm 0.007$&$13.673\pm 0.100$&$584.505$&$593.348$\\ \hline
\end{tabular}}
\end{center}
\end{sidewaystable}


\begin{sidewaystable}
\begin{center}
\caption{Results of the data fitting using the joint analysis from Union 2.1, $\rm{H(z)}$ and BAO. The error informed corresponds to 68$\%$ confidence level. Fixed means that the parameter was set to zero and the dashed lines mean that the model does not have that parameter. The derived parameters are: the current value of the deceleration parameter $q_0$, the value of the effective state parameter today $w_{\rm{eff}}$ and the calculated age of the universe in Gy. The AIC and BIC parameters are indicated in each case.
\label{Tablea2}}
\scalebox{0.65}{
\begin{tabular}{llllllllllll}
\\ \hline
Model&$\Omega_{\rm{m}0}$ & $\gamma_{\rm{x}}$ & $\alpha$ & $\beta$ &  $h$ & $q_0$ & $\omega_{\rm{eff}}$ & Age & AIC & BIC \\ \hline
$\Gamma_{1a}$ &$0.243\pm 0.026$&Fixed&$0.0064\pm 0.0112$&$-0.0300\pm 0.0699$&$0.699\pm 0.003$&$-0.567\pm 0.040$&$-0.711\pm 0.026$&$13.787\pm 0.687$&$587.790$&$605.470$\\ [1.0ex]
$\Gamma_{1b}$ &$0.247\pm 1.608$&$-0.059\pm 2.377$&Fixed&$0.0049\pm 2.4362$&$0.701\pm 0.004$&$-0.624\pm 3.591$&$-0.750\pm 2.394$&$13.614\pm 27.348$&$587.422$&$605.102$\\ [1.0ex]
$\Gamma_{1c}$ &$0.246\pm 0.044$&$-0.053\pm 0.134$&$0.0012\pm 0.0128$&Fixed&$0.701\pm 0.004$&$-0.618\pm 0.158$&$-0.746\pm 0.105$&$13.637\pm 0.655$&$587.398$&$605.078$\\ [1.0ex]
$\Gamma_{1d}$ &$0.246\pm 0.049$&$-0.053\pm 0.141$&$0.0010\pm 0.0127$&$0.0010\pm 0.0127$&$0.701\pm 0.004$&$-0.619\pm 0.169$&$-0.746\pm 0.112$&$13.633\pm 0.734$&$587.400$&$605.080$\\ [1.0ex]
$\Gamma_{1e}$ &$0.238\pm 0.022$&Fixed&Fixed&$0.0045\pm 0.0519$&$0.699\pm 0.003$&$-0.574\pm 0.033$&$-0.716\pm 0.022$&$13.654\pm 0.476$&$586.490$&$599.750$\\ [1.0ex]
$\Gamma_{1f}$ &$0.235\pm 0.015$&Fixed&$0.0037\pm 0.0094$&Fixed&$0.699\pm 0.003$&$-0.579\pm 0.022$&$-0.719\pm 0.015$&$13.687\pm 0.219$&$586.048$&$599.308$\\ [1.0ex]
$\Gamma_{1g}$ &$0.235\pm 0.016$&Fixed&$0.0032\pm 0.0086$&$0.0032\pm 0.0086$&$0.699\pm 0.003$&$-0.580\pm 0.024$&$-0.720\pm 0.016$&$13.676\pm 0.237$&$586.108$&$599.368$\\ [1.0ex]
$\Gamma_{2b}$ &$0.247\pm 1.695$&$-0.059\pm 2.505$&Fixed&$0.0735\pm 40.0856$&$0.701\pm 0.004$&$-0.624\pm 3.784$&$-0.749\pm 2.523$&$13.615\pm 29.426$&$587.422$&$605.102$\\ [1.0ex]
$\Gamma_{2c}$ &$0.246\pm 0.044$&$-0.053\pm 0.134$&$-0.0012\pm 0.0128$&Fixed&$0.701\pm 0.004$&$-0.619\pm 0.158$&$-0.746\pm 0.105$&$13.636\pm 0.654$&$587.398$&$605.078$\\ [1.0ex]
$\Gamma_{2d}$ &$0.246\pm 0.043$&$-0.052\pm 0.134$&$-0.0012\pm 0.0128$&$-0.0012\pm 0.0128$&$0.701\pm 0.004$&$-0.618\pm 0.158$&$-0.746\pm 0.105$&$13.637\pm 0.654$&$587.398$&$605.078$\\ [1.0ex]
$\Gamma_{2f}$ &$0.235\pm 0.015$&Fixed&$-0.0037\pm 0.0094$&Fixed&$0.699\pm 0.003$&$-0.579\pm 0.022$&$-0.719\pm 0.015$&$13.687\pm 0.219$&$586.048$&$599.308$\\ [1.0ex]
$\Gamma_{3}$ &$0.246\pm 0.043$&$-0.052\pm 0.133$&$0.0007\pm 0.0071$&$---$&$0.701\pm 0.004$&$-0.618\pm 0.156$&$-0.746\pm 0.104$&$13.638\pm 0.640$&$587.398$&$605.078$\\ [1.0ex]
$\Gamma_{3a}$ &$0.235\pm 0.014$&Fixed&$0.0019\pm 0.0048$&$---$&$0.699\pm 0.003$&$-0.579\pm 0.022$&$-0.719\pm 0.014$&$13.690\pm 0.213$&$586.040$&$599.300$\\ [1.0ex]
$\Gamma_{4}$ &$0.246\pm 0.042$&$-0.052\pm 0.132$&$0.0012\pm 0.0130$&$---$&$0.701\pm 0.004$&$-0.618\pm 0.155$&$-0.745\pm 0.103$&$13.638\pm 0.626$&$587.398$&$605.078$\\ [1.0ex]
$\Gamma_{4a}$ &$0.235\pm 0.014$&Fixed&$0.0040\pm 0.0098$&$---$&$0.699\pm 0.003$&$-0.579\pm 0.021$&$-0.719\pm 0.014$&$13.694\pm 0.207$&$586.031$&$599.291$\\ [1.0ex]
$\Gamma_{5}$ &$0.248\pm 0.162$&$-0.060\pm 0.215$&$0.0060\pm 0.4401$&$---$&$0.701\pm 0.004$&$-0.624\pm 0.344$&$-0.749\pm 0.230$&$13.618\pm 2.783$&$587.428$&$605.108$\\ [1.0ex]
$\Gamma_{5a}$ &$0.241\pm 0.023$&Fixed&$-0.0050\pm 0.0888$&$---$&$0.699\pm 0.003$&$-0.570\pm 0.035$&$-0.713\pm 0.023$&$13.673\pm 0.497$&$586.497$&$599.757$\\ [1.0ex]
$\Gamma_{6}$ &$0.246\pm 0.047$&$-0.053\pm 0.137$&$0.0008\pm 0.0097$&$---$&$0.701\pm 0.004$&$-0.619\pm 0.164$&$-0.746\pm 0.109$&$13.634\pm 0.712$&$587.400$&$605.080$\\ [1.0ex]
$\Gamma_{6a}$ &$0.235\pm 0.015$&Fixed&$0.0025\pm 0.0068$&$---$&$0.699\pm 0.003$&$-0.580\pm 0.023$&$-0.720\pm 0.015$&$13.680\pm 0.243$&$586.093$&$599.353$\\ [1.0ex]
$\Gamma_{7}$ &$0.246\pm 0.043$&$-0.052\pm 0.133$&$-0.0008\pm 0.0095$&$---$&$0.701\pm 0.004$&$-0.618\pm 0.157$&$-0.746\pm 0.105$&$13.638\pm 0.645$&$587.398$&$605.078$\\ [1.0ex]
$\Gamma_{7a}$ &$0.235\pm 0.015$&Fixed&$-0.0028\pm 0.0071$&$---$&$0.699\pm 0.003$&$-0.579\pm 0.022$&$-0.719\pm 0.015$&$13.690\pm 0.216$&$586.046$&$599.306$\\ [1.0ex]
$\Gamma_{8}$ &$0.247\pm 0.035$&$-0.051\pm 0.123$&$0.0020\pm 0.0180$&$---$&$0.701\pm 0.004$&$-0.616\pm 0.142$&$-0.744\pm 0.095$&$13.646\pm 0.542$&$587.396$&$605.076$\\ [1.0ex]
$\Gamma_{8a}$ &$0.237\pm 0.013$&Fixed&$0.0065\pm 0.0148$&$---$&$0.699\pm 0.003$&$-0.577\pm 0.020$&$-0.718\pm 0.013$&$13.715\pm 0.195$&$585.950$&$599.210$\\ [1.0ex]
$\omega$CDM&$0.249\pm 0.027$&$-0.059\pm 0.093$&$---$&$---$&$0.701\pm 0.004$&$-0.621\pm 0.108$&$-0.748\pm 0.072$&$13.626\pm 0.409$&$585.433$&$598.693$\\ [1.0ex]
$\Lambda$CDM&$0.240\pm 0.014$&$---$&$---$&$---$&$0.699\pm 0.003$&$-0.572\pm 0.021$&$-0.714\pm 0.014$&$13.666\pm 0.189$&$584.502$&$593.342$\\ \hline
\end{tabular}}
\end{center}
\end{sidewaystable}


\begin{sidewaystable}
\begin{center}
\caption{Results of the data fitting using the joint analysis from Union 2.1 and $\rm{H(z)}$.
\label{Tableb}}
\scalebox{0.65}{
\begin{tabular}{llllllllllll}
 \hline
Model &$\Omega_{\rm{m}0}$ & $\gamma_{\rm{x}}$ & $\alpha$ & $\beta$ &  $h$ & $q_0$ & $\omega_{\rm{eff}}$ & Age & AIC & BIC \\ \hline
$\Gamma_{1a}$ &$0.238\pm 0.081$&Fixed&$0.1853\pm 0.4531$&$-0.1207\pm 0.4354$&$0.700\pm 0.005$&$-0.574\pm 0.121$&$-0.716\pm 0.081$&$13.050\pm 5.700$&$585.800$&$603.440$\\ [1.0ex]
$\Gamma_{1c}$ &$0.159\pm 0.196$&$0.100\pm 0.315$&$0.2059\pm 0.5726$&Fixed&$0.700\pm 0.005$&$-0.574\pm 0.460$&$-0.716\pm 0.307$&$13.050\pm 5.214$&$585.800$&$603.440$\\ [1.0ex]
$\Gamma_{1d}$ &$0.010\pm 0.472$&$0.236\pm 0.418$&$0.1954\pm 0.3892$&$0.1954\pm 0.3892$&$0.700\pm 0.004$&$-0.582\pm 0.802$&$-0.722\pm 0.534$&$13.163\pm 7.475$&$585.807$&$603.448$\\ [1.0ex]
$\Gamma_{1e}$ &$0.212\pm 0.055$&Fixed&Fixed&$0.0467\pm 0.1294$&$0.701\pm 0.004$&$-0.613\pm 0.082$&$-0.742\pm 0.055$&$13.650\pm 1.149$&$583.985$&$597.215$\\ [1.0ex]
$\Gamma_{1f}$ &$0.217\pm 0.032$&Fixed&$0.0650\pm 0.1294$&Fixed&$0.701\pm 0.004$&$-0.607\pm 0.048$&$-0.738\pm 0.032$&$13.415\pm 1.129$&$583.888$&$597.119$\\ [1.0ex]
$\Gamma_{1g}$ &$0.213\pm 0.044$&Fixed&$0.0285\pm 0.0656$&$0.0285\pm 0.0656$&$0.701\pm 0.004$&$-0.612\pm 0.066$&$-0.741\pm 0.044$&$13.542\pm 0.911$&$583.937$&$597.168$\\ [1.0ex]
$\Gamma_{2c}$ &$0.159\pm 0.196$&$0.100\pm 0.315$&$-0.1709\pm 0.3938$&Fixed&$0.700\pm 0.005$&$-0.574\pm 0.460$&$-0.716\pm 0.307$&$13.049\pm 5.216$&$585.800$&$603.440$\\ [1.0ex]
$\Gamma_{2d}$ &$0.140\pm 0.321$&$0.120\pm 0.436$&$-0.2102\pm 0.6193$&$-0.2102\pm 0.6193$&$0.700\pm 0.005$&$-0.574\pm 0.680$&$-0.716\pm 0.454$&$13.051\pm 7.578$&$585.800$&$603.440$\\ [1.0ex]
$\Gamma_{2f}$ &$0.217\pm 0.032$&Fixed&$-0.0610\pm 0.1141$&Fixed&$0.701\pm 0.004$&$-0.607\pm 0.048$&$-0.738\pm 0.032$&$13.415\pm 1.128$&$583.888$&$597.119$\\ [1.0ex]
$\Gamma_{3}$ &$0.175\pm 0.194$&$0.079\pm 0.323$&$0.0888\pm 0.2004$&$---$&$0.700\pm 0.005$&$-0.578\pm 0.464$&$-0.719\pm 0.309$&$13.025\pm 4.765$&$585.817$&$603.457$\\ [1.0ex]
$\Gamma_{3a}$ &$0.218\pm 0.029$&Fixed&$0.0426\pm 0.0823$&$---$&$0.701\pm 0.004$&$-0.605\pm 0.044$&$-0.737\pm 0.029$&$13.355\pm 1.181$&$583.880$&$597.110$\\ [1.0ex]
$\Gamma_{4}$ &$0.203\pm 0.125$&$0.039\pm 0.254$&$0.2036\pm 0.6699$&$---$&$0.700\pm 0.005$&$-0.584\pm 0.338$&$-0.723\pm 0.226$&$13.012\pm 4.151$&$585.841$&$603.481$\\ [1.0ex]
$\Gamma_{4a}$ &$0.221\pm 0.025$&Fixed&$0.1208\pm 0.2244$&$---$&$0.701\pm 0.004$&$-0.601\pm 0.037$&$-0.734\pm 0.025$&$13.247\pm 1.287$&$583.868$&$597.098$\\ [1.0ex]
$\Gamma_{5a}$ &$0.213\pm 0.065$&Fixed&$0.0684\pm 0.2297$&$---$&$0.701\pm 0.004$&$-0.613\pm 0.098$&$-0.742\pm 0.065$&$13.684\pm 1.263$&$584.025$&$597.256$\\ [1.0ex]
$\Gamma_{6}$ &$0.010\pm 0.292$&$0.238\pm 0.820$&$0.1805\pm 0.6988$&$---$&$0.700\pm 0.005$&$-0.579\pm 1.209$&$-0.719\pm 0.806$&$13.137\pm 10.823$&$585.800$&$603.441$\\ [1.0ex]
$\Gamma_{6a}$ &$0.214\pm 0.043$&Fixed&$0.0263\pm 0.0599$&$---$&$0.701\pm 0.004$&$-0.612\pm 0.065$&$-0.741\pm 0.043$&$13.534\pm 1.133$&$583.933$&$597.164$\\ [1.0ex]
$\Gamma_{7}$ &$0.136\pm 0.326$&$0.126\pm 0.439$&$-0.1828\pm 0.5264$&$---$&$0.700\pm 0.005$&$-0.573\pm 0.688$&$-0.715\pm 0.459$&$13.059\pm 6.989$&$585.796$&$603.437$\\ [1.0ex]
$\Gamma_{7a}$ &$0.217\pm 0.032$&Fixed&$-0.0544\pm 0.1089$&$---$&$0.701\pm 0.004$&$-0.607\pm 0.049$&$-0.738\pm 0.032$&$13.420\pm 1.106$&$583.889$&$597.119$\\ [1.0ex]
$\Gamma_{8}$ &$0.355\pm 0.228$&$-0.193\pm 0.362$&$0.2676\pm 0.5653$&$---$&$0.700\pm 0.005$&$-0.573\pm 0.523$&$-0.715\pm 0.349$&$13.059\pm 5.438$&$585.796$&$603.437$\\ [1.0ex]
$\Gamma_{8a}$ &$0.218\pm 0.076$&Fixed&$-0.0508\pm 0.2731$&$---$&$0.701\pm 0.005$&$-0.606\pm 0.114$&$-0.737\pm 0.076$&$13.800\pm 1.166$&$584.077$&$597.308$\\ [1.0ex]
$\omega$CDM &$0.246\pm 0.041$&$-0.047\pm 0.129$&$---$&$---$&$0.701\pm 0.004$&$-0.613\pm 0.152$&$-0.742\pm 0.101$&$13.650\pm 0.607$&$583.985$&$597.215$\\ [1.0ex]
$\Lambda$CDM &$0.231\pm 0.016$&$---$&$---$&$---$&$0.700\pm 0.003$&$-0.585\pm 0.023$&$-0.723\pm 0.016$&$13.758\pm 0.221$&$582.118$&$590.938$\\ \hline
\end{tabular}}
\end{center}
\end{sidewaystable}


\begin{table}[ht!]
\centering
\caption{Ranking of models according to BIC. In the left panel we show the joint analysis of Union 2.1+$\rm{H(z)}$+BAO+CMB and in the right panel we have the joint analysis of binned JLA+$\rm{H(z)}+$BAO+CMB as comparison. f.p. is the number of free parameters in the model.\label{Tablec}}
\begin{tabular*}{\textwidth}{@{\extracolsep{\fill}}l l l l@{}}
 \\ \hline
Model        &{U2.1+BAO+\rm{H(z)}+CMB}&{bJLA+BAO+\rm{H(z)}+CMB}&f.p.\\ \hline
$\Lambda$CDM  &593.348&58.972&2\\ [0.6ex] 
$\omega$CDM   &598.700&62.457&3\\ [0.6ex]
$\Gamma_{8a}$ &599.362&62.346&3\\ [0.6ex]
$\Gamma_{7a}$ &599.367&62.389&3\\ [0.6ex]
$\Gamma_{6a}$ &599.384&62.429&3\\ [0.6ex]
$\Gamma_{1g}$ &599.726&62.543&3\\ [0.6ex]
$\Gamma_{1f}$ &599.729&62.540&3\\ [0.6ex]
$\Gamma_{2f}$ &599.732&62.860&3\\ [0.6ex]
$\Gamma_{4a}$ &599.741&62.544&3\\ [0.6ex]
$\Gamma_{1e}$ &599.767&63.087&3\\ [0.6ex]
$\Gamma_{5a}$ &599.885&63.340&3\\ [0.6ex]
$\Gamma_{3a}$ &599.890&62.542&3\\ [0.6ex]
$\Gamma_{7}$  &605.107&66.359&4\\ [0.6ex]
$\Gamma_{1b}$ &605.108&66.619&4\\ [0.6ex]
$\Gamma_{2b}$ &605.120&66.647&4\\ [0.6ex]
$\Gamma_{1d}$ &605.122&66.358&4\\ [0.6ex]
$\Gamma_{1c}$ &605.123&66.357&4\\ [0.6ex]
$\Gamma_{6}$  &605.126&66.394&4\\ [0.6ex]
$\Gamma_{5}$  &605.135&66.953&4\\ [0.6ex]
$\Gamma_{2c}$ &605.143&66.358&4\\ [0.6ex]
$\Gamma_{2d}$ &605.175&66.482&4\\ [0.6ex]
$\Gamma_{4}$  &605.379&66.358&4\\ [0.6ex]
$\Gamma_{1a}$ &606.137&66.728&4\\ [0.6ex]
$\Gamma_{3}$  &606.009&66.433&4\\ [0.6ex]
$\Gamma_{8}$  &606.991&67.219&4\\ \hline

\end{tabular*}
\end{table}


In Table \ref{Tablea} the best fit parameters for all the analyzed models are shown; we used a joint analysis considering Union 2.1+$\rm{H(z)}$+BAO+CMB. The subscripts $a$, $b$, $c$, $d$, $e$, $f$, $g$  in the models denote $\gamma_{\rm{x}}=0$, $\alpha=0$, $\beta=0$, $\alpha=\beta$, $\gamma_{\rm{x}}=\alpha=0$, $\gamma_{\rm{x}}=\beta=0$ and $\gamma_{\rm{x}}=0$ with $\alpha=\beta$, respectively. From Table \ref{T1} and in the context of this classification we note that $\Gamma_{2e}$ does not correspond to an interacting model, because the parameters $b_1$, $b_2$ and $b_3$ in Table \ref{T1} have fixed values in this case. Because of this, $\Gamma_{2e}$ is not present in Tables \ref{Tablea} - \ref{Tablec}. Also, we note that the only difference between $\Gamma_{1f}$ and $\Gamma_{2g}$ is a sign in the interaction term, thus we exclude $\Gamma_{2g}$ from the analysis.

In Table \ref{Tablea} we have also included, besides interacting models, $\Lambda$CDM and $\omega$CDM models as comparison. 
In this table all interacting scenarios and $\omega$CDM model present a negative value of the barotropic index of DE ($\gamma_x$), indicating that there is a trend in favor of phantom DE models. Nevertheless, $\gamma_{\rm{x}}$ is compatible with zero considering the 1$\sigma$ confidence level. Besides, we note that some of the interacting parameters become smaller than $5\times10^{-5}$ when we include CMB data in the analysis, this is the case for $\Gamma_{1c}$ and $\Gamma_{1d}$.  Also, we note that interaction $\Gamma_{2a}$ is not well constrained by the considered data and some of the interactions have a defined sign inside the 1$\sigma$ region, this is the case of $\Gamma_{1b}$, $\Gamma_{1e}$, $\Gamma_{1f}$, $\Gamma_{2f}$, $\Gamma_{3a}$, $\Gamma_{4a}$, $\Gamma_{5}$, $\Gamma_{5a},\ \Gamma_{6a},\ \Gamma_{7a}$ and $\Gamma_{8a}$.

In Table \ref{Tablea2} we show the joint analysis considering only Union 2.1+$\rm{H(z)}$+BAO, we note that the case $\Gamma_{2a}$ is absent because the error in the $\beta$ parameter becomes too large (which we can also observe in Table \ref{Tablea}). Here, $\gamma_x$ is negative in all the cases and most of the interacting models have the same sign in the interacting parameters as in Table \ref{Tablea}, but $\Gamma_{1a},\ \Gamma_{2d},\ \Gamma_{5},\ \Gamma_{5a}$. Also, in comparing Table \ref{Tablea2} to Table \ref{Tablea} we note that interactions $\Gamma_{1b}$, $\Gamma_{1e}$, $\Gamma_{5}$, $\Gamma_{6a}$, $\Gamma_{7a}$, $\Gamma_{8}$ and $\Gamma_{8a}$ have the same order of magnitude for interacting parameters when we include CMB data. Interactions $\Gamma_{5a}$, $\Gamma_{6}$ and $\Gamma_{7}$ increase the values of the interacting parameter and the remaining cases reduce their absolute value in one or two orders of magnitude when we consider CMB data.

In Table \ref{Tableb} we show the joint analysis considering only Union 2.1 and $\rm{H(z)}$ data.
We note that most of interactions have $\gamma_x>0$, indicating that it is BAO and CMB data which constrain this parameter to be negative. On the other hand, we do not include in this table interactions $\Gamma_{1b}$, $\Gamma_{2a}$, $\Gamma_{2b}$ and $\Gamma_{5}$ because the error in the interaction parameters in these cases become too large, as we can see in Table \ref{Tablea2} for $\Gamma_{1b}$, $\Gamma_{2b}$ and $\Gamma_{5}$ and in Table \ref{Tablea} for $\Gamma_{2a}$. 

In Tables \ref{Tablea} - \ref{Tableb} we notice that, even though there is a deviation from the $\Lambda$CDM scenario, we obtain similar values for the current deceleration parameter $q_0$, the current effective state parameter $\omega_{\rm{eff}}$ and the age of our universe for all the studied interacting scenarios.

In Table \ref{Tablec} we extend our analysis by considering binned data of the more recent JLA compilation of SN Ia \cite{JLA}. We note that for the joint analysis using Union 2.1 or JLA compilation the results are consistent, and in light of the Bayesian information criterion, the interacting models are ordered according to the number of free parameters of each model.

\begin{figure}[ht!]
\centering
\includegraphics[width=0.9\textwidth]{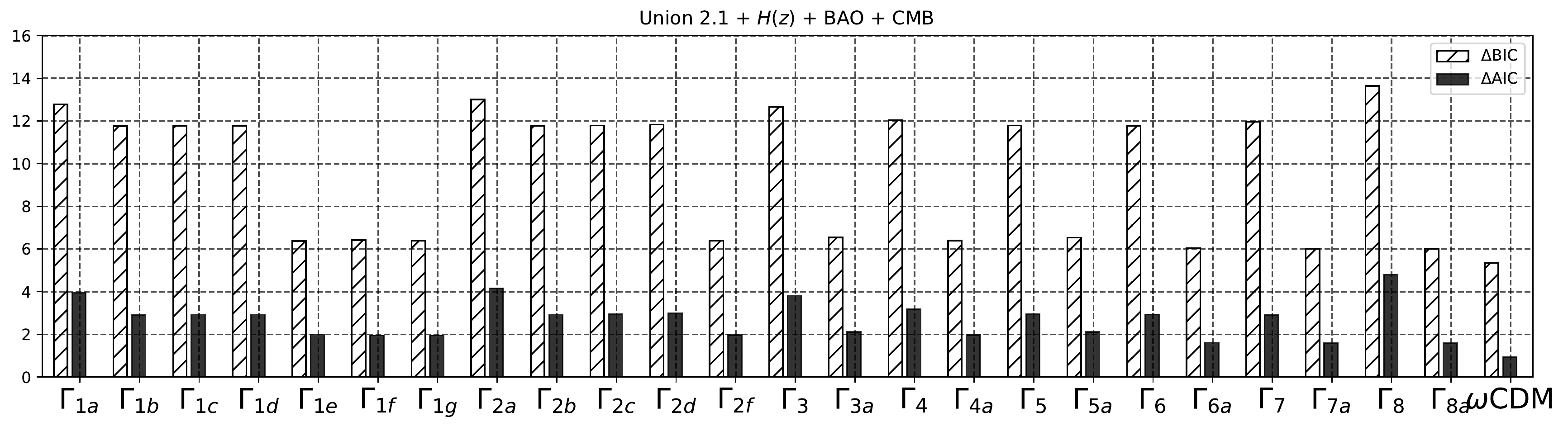}
\caption{$\Delta$AIC and $\Delta$BIC of models defined in Table \ref{Tablea} compared to $\Lambda$CDM.}
\label{graphic1}
\end{figure}

\begin{figure}[ht!]
\centering
\includegraphics[width=0.9\textwidth]{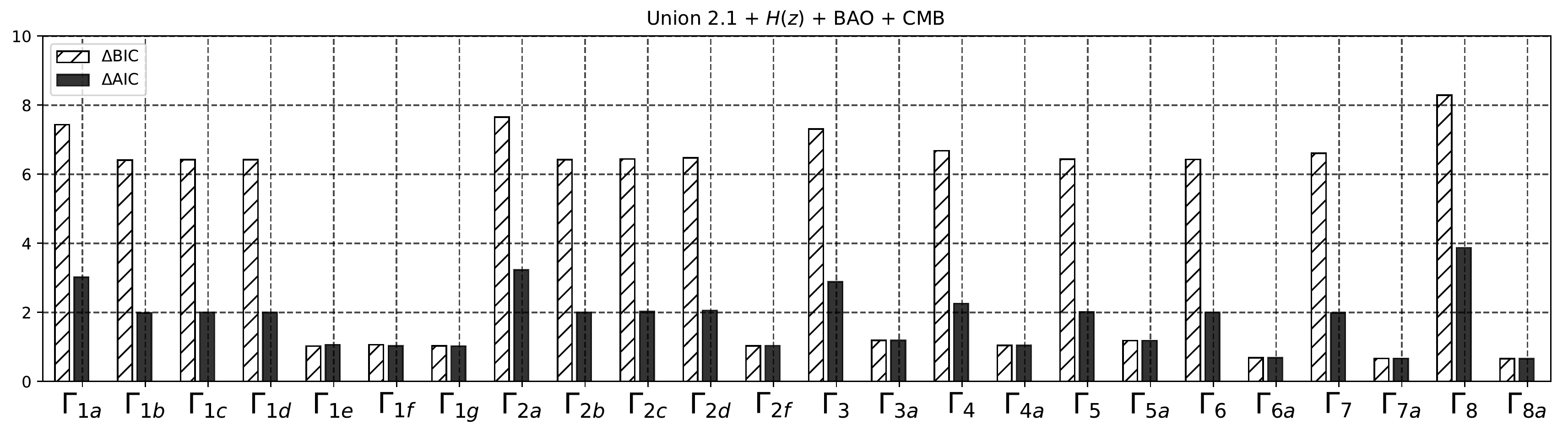}
\caption{$\Delta$AIC and $\Delta$BIC of interacting models defined in Table \ref{Tablea} compared to the $\omega$CDM model.}
\label{graphic2}  
\end{figure}

In our analysis $\Lambda$CDM is the model with the lowest AIC and BIC parameters when we use data from the joint analysis of Union2.1+$\rm{H(z)}$+BAO+CMB (Table \ref{Tablea}), Union2.1+$\rm{H(z)}$+BAO (Table \ref{Tablea2}), Union2.1+$\rm{H(z)}$ (Table \ref{Tableb}) or binned JLA+$\rm{H(z)}$+BAO+CMB (Table \ref{Tablec}). From Figure \ref{graphic1} we see that, when the underlying model is assumed to be $\Lambda$CDM, AIC indicates that all models with three free parameters are in the region of ``strong evidence in favor''. Nevertheless under BIC, interacting models with four free parameters are further than having ``strong evidence against'' and the models of three free parameters are in the upper limit of having ``evidence against''. From Figures \ref{graphic1} and \ref{graphic2}, we notice a tension between AIC and BIC results, while AIC indicates there is ``evidence in favor'' BIC indicates that there is ``evidence against'' or ``strong evidence against'' for the same model. This is due to the fact that BIC strongly penalizes models when they have a larger number of parameters \cite{Liddle2}. 

Compared to $\Lambda$CDM, the studied interacting models have ``evidence against''.
This is consistent with the results of Ref.\cite{baye2}, where the authors conclude that the particular interacting model they study is disfavored compared to $\Lambda$CDM, also they notice that BIC is a more restrictive criteria.  The model $\omega$CDM is also incompatible with $\Lambda$CDM with respect to BIC. 

\begin{figure}
\centering
\includegraphics[width=0.65\textwidth]{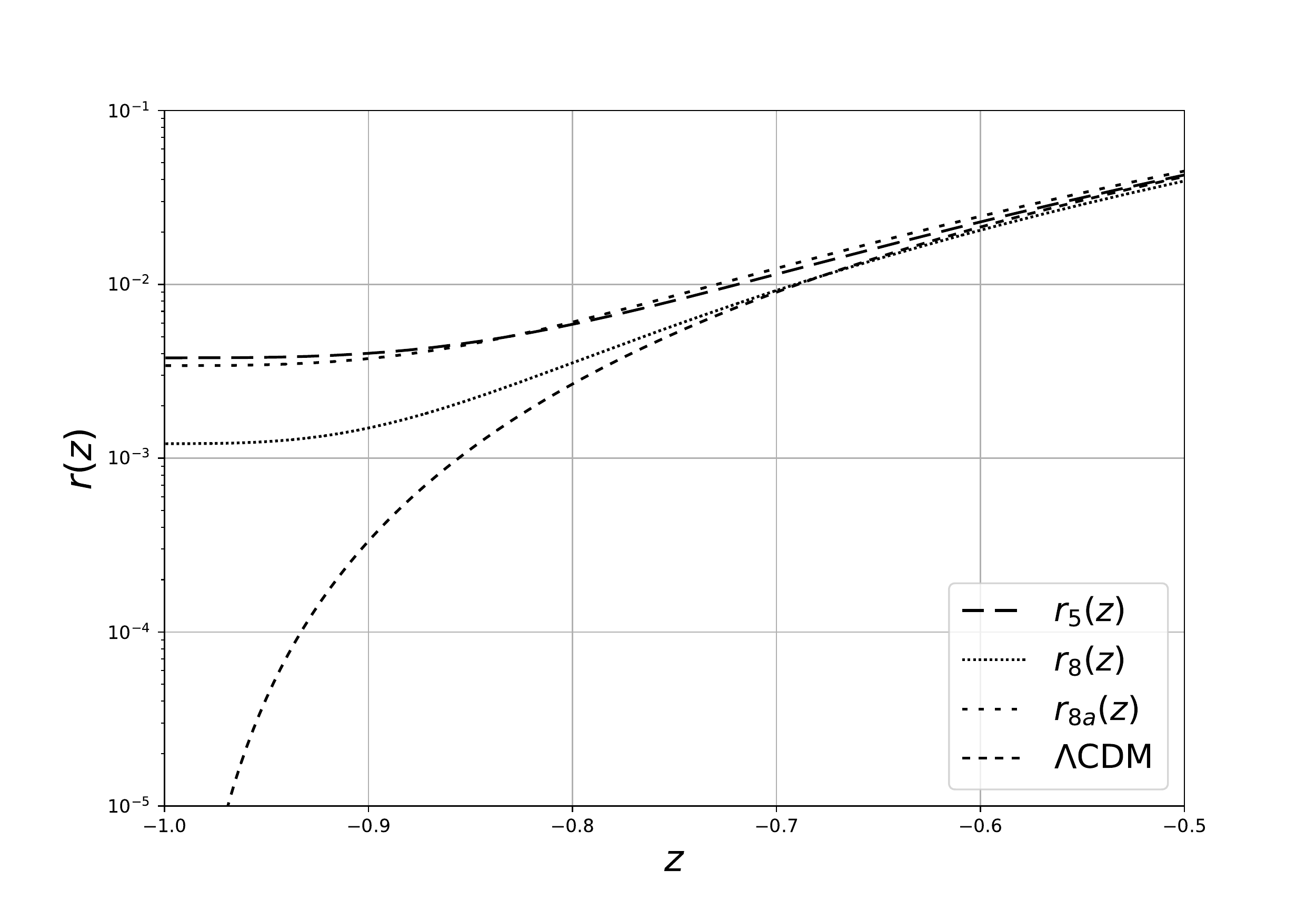}
\caption{\label{r(z)} Coincidence parameter in semilog scale. These interactions have an energy transfer from DE to DM.}
\end{figure}

\begin{figure}
\centering
\includegraphics[width=0.65\textwidth]{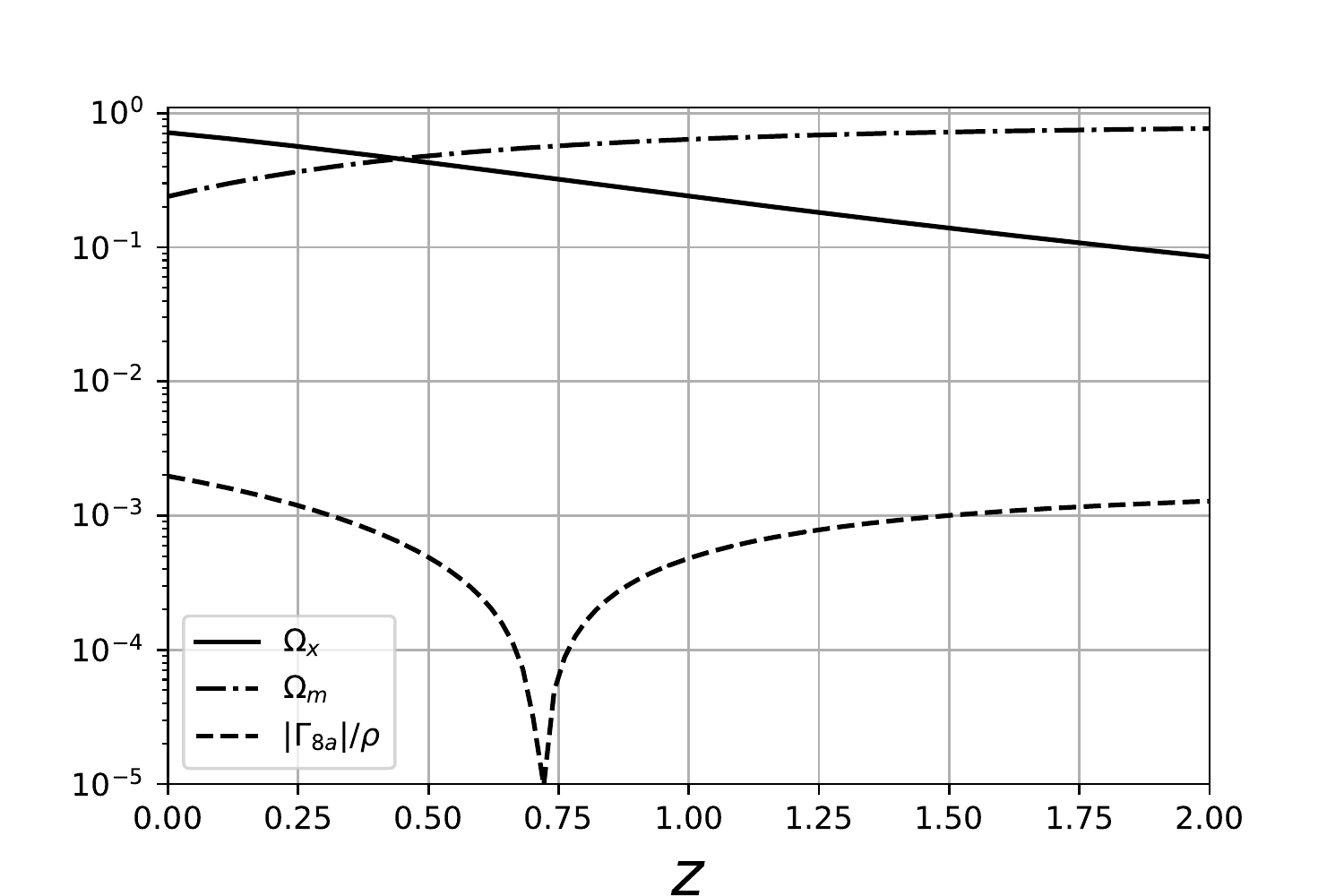}
\caption{\label{gamma8} Semilog graphic of the evolution of the density parameters for the interacting model $\Gamma_{8a}$, note that the interaction has a sign change at redshift $z\approx0.7$ approximately.}
\end{figure}

If we compare the models without considering $\Lambda$CDM, the best model according to AIC and BIC is $\omega$CDM when we consider the joint analysis of Union2.1+$\rm{H(z)}$+ BAO+CMB. In Table \ref{Tablec} we consider only the more stringent criteria, BIC. Here we note that under BIC all models with three free parameters (f.p.) cannot be ruled out when we assume that $\omega$CDM is the underlying model. In Figure \ref{graphic2} we see that by using BIC there is ``strong evidence against'' models with 4 f.p. when the base model is $\omega$CDM, i.e., we can rule out models of 4 f.p. but not models of 3 f.p. if the best model is $\omega$CDM. On the other hand, the best interacting model under BIC (and AIC) is $\Gamma_{8a}$, which has an interaction proportional to the deceleration parameter $q$.
Among all our models, those shown in Figure \ref{r(z)} alleviate the coincidence problem, besides, all of them have an energy transfer from DE to DM today. In the case of $\Gamma_{8a}$, for $z\gtrsim0.7$ we have an energy transfer from DM to DE and for $z\lesssim 0.7$ the energy transfer is from DE to DM as we see in Figure \ref{gamma8}.  

It is noteworthy to mention that interaction $\Gamma_{8a}$ is marginally better than other interacting models according to AIC and BIC and this interaction alleviates the coincidence problem and changes sign during evolution. A similar behavior was reported in Ref.\cite{cai} where the authors separate the data in redshift bins for $Q=3H\delta$, where $\delta$ is a constant fitted for each bin. The authors consider different parametrizations of the equation of state for DE and they found an oscillation of the interaction sign. Sign-changeable interactions have also been studied in Refs. \cite{wei}-\cite{schg}, \cite{Fabiola}.

As summary, from our analysis we notice that there are consistent interacting models that explain the data equally well than $\omega$CDM, and an increase of the number of free parameters in interacting models, although phenomenologically interesting, is strongly penalized according to BIC in the description of the late universe.

%
%



\section{Final Remarks}
\label{FR}

In this work we analyzed eight general types of interacting models of the dark sector with analytical solutions and compared how well they fit the joint data from Union 2.1+$\rm{H(z)}$+BAO+CMB using the Akaike information criterion and the Bayesian information criterion. The main goal of our work was to investigate if more complex interacting models (more complex meaning models with more free parameters) are competitive in fitting the data and whether we could distinguish them via AIC and BIC.

The models in Table \ref{T1} are interesting because they are good candidates to alleviate the coincidence problem, furthermore, the physical motivation to the studied models was discussed in section \ref{IM}, where we showed that the family of interactions presented can be interpreted in terms of a variable Chaplygin gas in a unified dark sector scenario or in terms of a variable state parameter for the dark energy component. 
  
Taking into account the theoretical problems that the $\Lambda$CDM scenario presents and the observational tensions recently reported with this model \cite{tension}, we assume that a departure from the simplest model is needed. We compared a family of interacting models among themselves and with the $\omega$CDM scenario. In our analysis we noted a tension between the results using AIC and BIC and we decided to follow the more stringent criterion, namely the BIC (Table \ref{Tablec}). According to our results, under the BIC ``there is not enough  evidence against'' any interacting model with three free parameters when we assume that the underlying model is the one which has the lowest BIC parameter, which turns out to be $\omega$CDM. Among the interacting models, $\Gamma_{8a}$ is the model with the lowest BIC parameter value, it corresponds to a sign-changeable interaction  with $\gamma_{\rm{x}}=0$ and $\gamma_{\rm{m}}=1$ and it is compatible with $\omega$CDM.
Furthermore, $\Gamma_{8a}$ is one of the models that alleviate the coincidence problem, since the value of the coincidence parameter in the future tends to a constant (see Fig. \ref{r(z)}). 

For the selected models we concluded that all the considered models with three free parameters are compatible among them, i.e. all they have a BIC parameter in the same range, thus these models are not distinguishable, generating in this sense a new kind of degeneracy problem. A similar behavior appears when we inspect models with four free parameters as we see in Table \ref{Tablec}. Furthermore, it is worth to emphasize that all the interacting models with three free parameters, besides of representing different phenomenology, adjust the data as well as the $\omega$CDM model.

When we compare models with three free parameters to models with four free parameters (using BIC) we find ``evidence against'' the four free parameters models when we assume that the underlying model is a three free parameters interacting model.

Finally we conclude that an increase of the complexity of interacting models, measured through the number of free parameters, is strongly penalized according to BIC in the description of the late universe. In the near future we expect to improve this analysis by considering different parametrizations for the DE state parameter, the dark degeneracy and more sophisticated methods to constrain data, such as Monte Carlo.

\section*{Acknowledgements}
This work was partially supported by Direcci\'on de Investigaci\'on de la Universidad del B\'io-B\'io through grants GI 150407/VC and 151307 3/R. F.A. has been supported by Comisi\'on Nacional de Ciencias y Tecnolog\'ia through Fondecyt Grant 3130736 and the Direcci\'on de Investigaci\'on de la Universidad de La Frontera, Project DI17-0075. We would like to thank Saulo Carneiro for the enlightening comments on the manuscript and Guillermo Rubilar for helpful discussions and the reviewing of the manuscript.


\renewcommand{\theequation}{A\arabic{equation}}
\setcounter{equation}{0}
\appendix
\section*{Appendix A:}

The distance modulus is defined as
\begin{equation}
\label{mu}
\mu(z):= 5\log\left[ \frac{d_L(z)}{1\rm{pc}}\right] -5,
\end{equation}
where $d_L(z)$ is the luminosity distance at redshift $z$. For a spatially flat universe, we have
\begin{equation}
d_L(z)=(1+z)r(z)=\frac{(1+z)c}{H_0}\int^z_0\frac{dz^\prime}{E(z^\prime)},
\end{equation}
with $H_0E(z)=H(z)$, $r(z)$ is the comoving radius at redshift $z$ and $c$ the speed of light.

On the other hand, the $\rm{H(z)}$ dataset is related to the measure of the age difference, $\Delta t$, between two passively evolving galaxies that formed at the same time but separated by a small redshift interval $\Delta z$. One can infer the value of the derivative, ($dz/dt$), from the ratio ($\Delta z/\Delta t$) \cite{liao} and through the relation
\begin{equation}
\label{Hz}
H(z)=-\frac{1}{1+z}\frac{dz}{dt},
\end{equation}
infer the value of $H$ for a given $z$.

For BAO's dataset we need to define the acoustic parameter introduced by Eisenstein and the BAO typical scale $r_s(z_d)$, i.e. the comoving radius of the sound horizon at the drag epoch $z_d$, when photons and baryons decouple.

The acoustic parameter $A(z)$ is given by \cite{Eisen},
\begin{equation}
\label{acoustic}
A(z)=\frac{D_{\rm{V}}(z)\sqrt{\hat{\Omega}_{\rm{m0}}} H_0^2}{cz},
\end{equation}
with $\hat{\Omega}_{\rm{m0}}=\Omega_{\rm{m0}}+\Omega_{\rm{b0}}$ where the distance scale $D_{\rm{V}}$ is defined as
\begin{equation}
D_{\rm{V}}(z)=\frac{1}{H_0}\left[(1+z)^2 D_A^2(z) \frac{cz}{E(z)}\right]^{\frac{1}{3}},
\end{equation}
and $D_A(z)$ is the angular diameter distance,
\begin{equation}
D_A(z)=\frac{D_L(z)}{(1+z)^2},
\end{equation}
with $D_L(z)=H_0d_L$.

Other important function is the dimensionless distance ratio given by
\begin{equation}
\label{dr}
d_z(z)=\frac{r_s(z_d)}{D_{\rm{V}}(z)},
\end{equation}
where the sound horizon is defined as
\begin{equation}
r_s(z)=\int_{z}^{\infty}\frac{c_s(z) dz}{H(z)},
\end{equation}
and the sound speed in the photon-baryon fluid is
\begin{equation}\label{sound}
c_s=\frac{c}{\sqrt{3(1+\mathcal{R})}},
\end{equation}
where $\mathcal{R}:=3\rho_{\rm{b}}/4\rho_{\gamma}$, $\rho_{\rm{b}}=\rho_{\rm{b}0}(1+z)^3$ is the energy density of baryons and $\rho_{\gamma}=\rho_{\gamma 0}(1+z)^4$ is the energy density of photons of the CMB radiation \cite{wein}. We use $\Omega_{\gamma0}h^2=2.469\times10^{-5}$\cite{wein} and $\Omega_{\rm{b0}} h^2=0.0222$ \cite{Planck} where $\Omega_{\gamma0}=\frac{\rho_{\gamma 0}}{3H_0^2}$ is the normalized energy density of CMB photons today, $\Omega_{\rm{b0}}=\frac{\rho_{\rm{b} 0}}{3H_0^2}$ is the normalized baryonic energy density today and $h$ is the dimensionless Hubble parameter such that $H_0=100h$ km s$^{-1}$Mpc$^{-1}$.

For the redshift at the drag epoch $z_d$ we use the formula proposed by Eisenstein to fit numerical recombination results \cite{Eisenstein}:
\begin{equation}
z_d=\frac{1291(\hat{\Omega}_{\rm{m0}}h^2)^{0.251}}{1+0.659(\hat{\Omega}_{\rm{m0}}h^2)^{0.828}}[1+\rm{b}_1(\Omega_{\rm{b0}}h^2)^{\rm{b}_2}],
\end{equation}
where
\begin{eqnarray*}
\rm{b}_1&=&0.313(\hat{\Omega}_{\rm{m0}}h^2)^{-0.419}[1+0.607(\hat{\Omega}_{\rm{m0}}h^2)^{0.674}],\\
\rm{b}_2&=&0.238(\hat{\Omega}_{\rm{m0}}h^2)^{0.223}.
\end{eqnarray*}
From the CMB we use the position of the first peak of the CMB anisotropy spectrum $l_1$ \cite{Hu}:
\begin{eqnarray}
\label{l1}
l_1=l_A(1-\delta_1)\  \ \textrm{where}\ \ \delta_1=0.267\left(\frac{r}{0.3}\right)^{0.1},
\end{eqnarray}
with $r=\rho_{\rm{r}}/(\rho_{\rm{m}}+\rho_{\rm{b}})$ evaluated at the redshift of last scattering $z_{\rm{ls}}$ and the radiation density given by \cite{wein}:
\begin{eqnarray}
\rho_{\rm{r}}(z)=3H_0^2\Omega_{\gamma0}\left(1+\frac{7}{8}\left(\frac{4}{11}\right)^{4/3}N_{\rm{eff}}\right)(1+z)^4,
\end{eqnarray}
where we have considered the neutrinos' contribution with $N_{\rm{eff}}=3.04$ \cite{Planck}.

The acoustic scale $l_A$ is defined as
\begin{eqnarray}
l_A=\frac{\pi d_L(z_{\rm{ls}})}{(1+z_{\rm{ls}})r_s(z_{\rm{ls}})},
\end{eqnarray}
where the last scattering redshift is approximated by \cite{HuSugiyama}:
\begin{eqnarray*}
z_{\rm{ls}}=1048\left(1+0.00124(\Omega_{\rm{b0}}h^2)^{-0.738}\right)\left(1+{\rm g}_1(\hat{\Omega}_{\rm{m0}}h^2)^{\rm{g_2}}\right),
\end{eqnarray*}
with:
\begin{eqnarray*}
{\rm g_1}=\frac{0.0783(\Omega_{\rm{b0}}h^2)^{-0.238}}{1+39.5(\Omega_{\rm{b0}}h^2)^{0.763}},\ {\rm g_2}=\frac{0.560}{1+21.1(\Omega_{\rm{b0}}h^2)^{1.81}}.
\end{eqnarray*}
\end{document}